# MONTAGEM DE UM SISTEMA DE PROJEÇÃO DIGITAL PARA DOMO HEMISFÉRICO

**FILHO**, Valdinei Bueno L.; **ASSUNÇÃO**, Hildeu Ferreira da; **LIMA**, Thiago Oliveira; **MARTINS,** Alessandro.

Campus Jataí, Universidade Federal de Goiás, Jataí - GO 75800-000, Brasil

**Introdução**

Existem atualmente no Brasil cerca de 23 planetários fixos, segundo dados da associação brasileira de planetários [1]. Em várias regiões do Brasil existem também os chamados planetários móveis, que ultimamente tem se tornado uma alternativa mais barata de planetário em relação aos fixos, apesar de inferiores tecnologicamente. Esse trabalho se baseia na montagem de um sistema de projeção digital full-dome caseiro para planetários infláveis. Os projetores comumente usados no Brasil, nas cúpulas portáteis (cúpula inflável), são os chamados opto-mecânicos [2]. Tratam-se de cilindros perfurados de forma apropriada, que envolvendo uma fonte luminosa podem criar a ilusão de se ver estrelas em um anteparo semi-esférico. As cúpulas infláveis dotadas de sistemas de projeção opto-mecânicos, tornaram-se uma alternativa de ensino boa e módica para o país. Entretanto, com a evolução dos sistemas de projeção tem surgido uma nova geração de planetários, os planetários digitais [3]. Porém, os altos preços desses produtos tem tornado essa uma opção inviável. Uma das motivações desse trabalho foi a possibilidade de criar um sistema de projeção digital caseiro com baixos custos. A cúpula inflável de utilizada para a projeção das imagens foi fabricada no Brasil por uma empresa especializada na construção de produtos infláveis de entretenimento, o que reduziu consideravelmente seu custo de obtenção, comparado à importação. Uma outra vantagem é a possibilidade de obtenção de níveis de qualidade extremamente superiores, proporcionados pelo sistema digital, bem como sua versatilidade. Estes sistemas de projeção digital podem ser utilizados para ensino de astronomia e ciências afins.

No município de Jataí existe um Campus da Universidade Federal de Goiás

(UFG) e um Centro Federal de Educação Tecnológica (CEFET). Apesar do grande potencial econômico e da existência de instituições de ensino tecnológico e superior federais, a região carece de um centro de divulgação científica, como museus de ciência, planetários, observatórios, etc. Os centros de divulgação científicos mais próximos como o Planetário fixo da UFG encontra-se em Goiânia, com uma distância mínima de 310 km, o que dificulta atingir um grande público, não somente do sudoeste goiano, mas também do próprio Estado. Desse modo, a construção do projetor digital tem permitido a criação de uma sala ambientada, através de um planetário inflável, onde são explorados aspectos da Astronomia.

**Metodologia**

Na montagem do sistema utilizamos acessórios ópticos e eletrônicos de baixo custo. Adquirimos um projetor digital Datashow de alta resolução (2300 lumens) para a reprodução de imagens. Para permitir a adaptação da imagem à curvatura do domo (projeção full dome), de modo a permitir que se cubra quase toda sua superfície hemisférica, utilizamos uma lente tipo "Olho-de-Peixe" de 8mm, acoplado a uma lente objetiva de 50 mm e uma lente de foco de 20 mm. Entretanto a obtenção das imagens através do sistema "Olho-de-Peixe" não é fácil. Características como ângulo de projeção do aparelho, distâncias focais da lente e do projetor são fatores decisivos na construção desse sistema de projeção. Desse modo, todo sistema de lentes foi acoplado a um conjunto de dois cilindros coaxiais ocos de latão, que funciona como um dispositivo de modo a permitir o ajuste manual do foco das imagens projetadas (uma das peças cilíndricas é deslizante em relação a outra fixa no suporte). Este dispositivo é então acoplado à saída de imagens do projetor Datashow e todo sistema montado sobre um suporte de madeira com uma estrutura que permite movimentar a projeção sobre 3 graus de liberdade (ver Fig. 1).

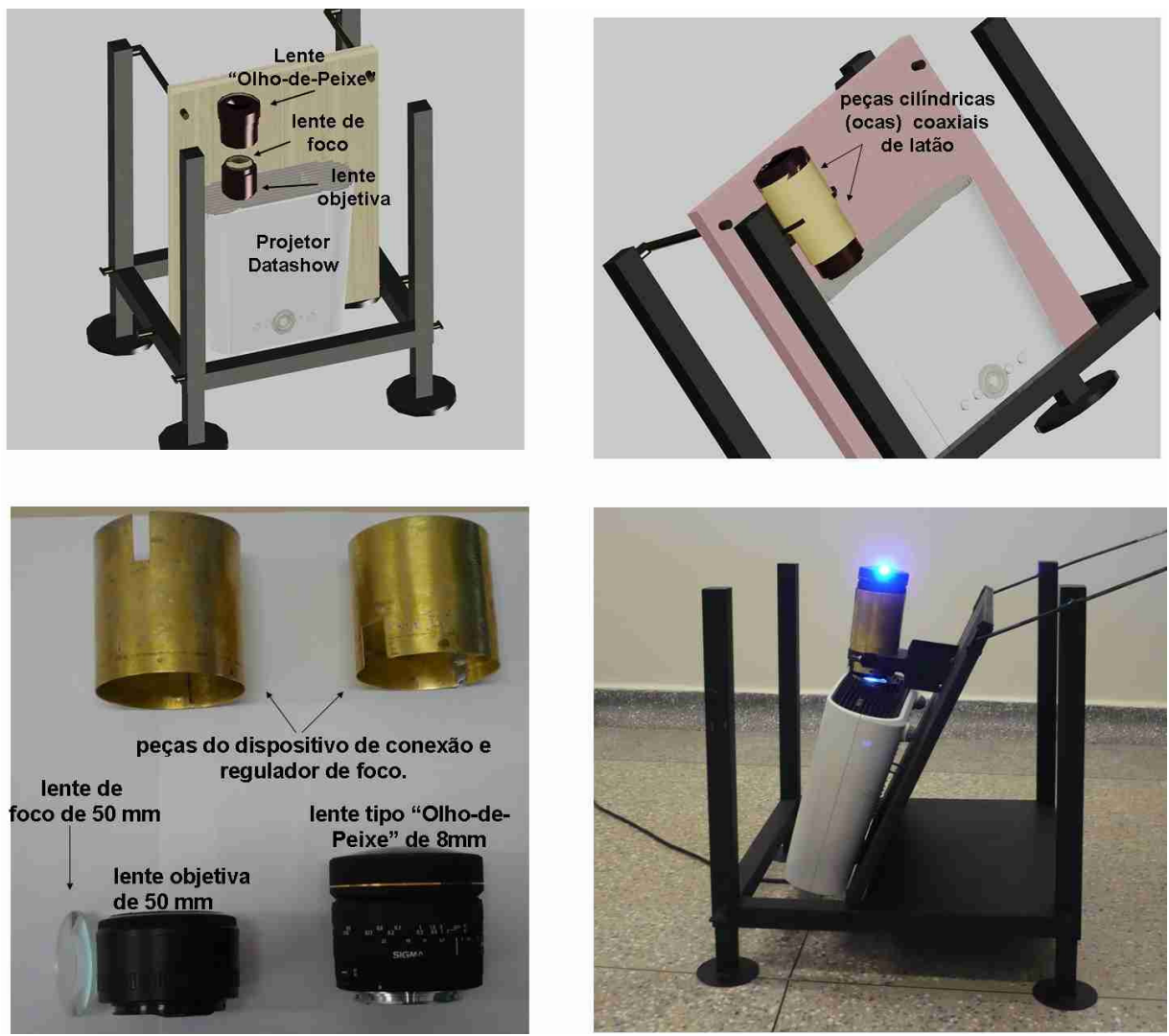

**Fig. 1 –** Imagens com a descrição da estrutura de montagem do sistema de projeção panorâmica digital (full-dome).

Conectado ao sistema de projeção de imagens panorâmico teremos um laptop, onde são instalados os softwares educativos usados para simular as imagens do planetário. Os softwares educativos utilizados são de acesso livre, gratuitos e de código-aberto que renderiza o céu em tempo real ou que utilizam imagens a partir de fotografias de satélite obtidas em fontes diversas [4-6].

**Resultados e discussão**

As sessões de planetário tem sido dinamizadas no Campus Jataí da UFG, em uma Tenda de 20 x 40 m². As visitações acontecem as segundas e terças-feiras, em período integral, onde as mesmas são previamente agendadas com as unidades de ensino da rede pública e particular, do município de Jataí e cidades vizinhas. Nestas sessões tem sido explorados aspectos como a origem do universo e formação da matéria, a formação e evolução do planeta Terra, a descrição de como se desenvolveram e evoluíram as técnicas de contagem do tempo e previsões climáticas, através da observação e do estudo do movimento dos corpos celestes.

**Conclusão**

A montagem do sistema de projeção digital tem permitido a reprodução de imagens com muito boa qualidade quando comparadas com as imagens características dos projetores industriais.

**Fonte financiadora**:

- Este projeto tem o apoio do Conselho Nacional de Desenvolvimento Científico e Tecnológico - CNPq

**Referências bibliográficas**